\documentstyle[12pt,epsfig]{ioplppt}
\def\beq{\begin{equation}}
\def\eeq{\end{equation}}
\begin{document}

\title{Interpretations  of $J/\psi$ suppression}

\author{N. Armesto and A. Capella}

\address{Laboratoire de Physique Th\'eorique et Hautes
Energies\footnote{Laboratoire associ\'e au Centre National de la Recherche
Scientifique - URA D0063.}\\ Universit\'e de Paris XI, B\^atiment 211, F-91405
Orsay Cedex, France}

\begin{abstract}
We review the two main interpretations of $J/\psi$ suppression proposed in the
literature.  The phase transition (or deconfining) scenario
assumes that below some critical value of the local energy density (or of some
other geometrical quantity which depends both on the colliding systems and on
the
centrality of the collision), there is only nuclear absorption. Above this
critical value the absorptive cross-section is taken to be infinite, i.e. no
$J/\psi$ can survive in this hot region. In the hadronic scenario the $J/\psi$
dissociates due both to nuclear absorption and to its interactions with
co-moving
hadrons produced in the collision. No discontinuity exists in physical
observables. We show that an equally good description of the
present data is possible in either scenario.

\end{abstract}

\section{Introduction} \par
\baselineskip=20 pt
\hspace*{\parindent} Charmonium suppression due to Debye screening in a
deconfined
medium was proposed in 1986 by Matsui and Satz \cite{1r} and found
experimentally
by the NA38 collaboration \cite{2r}. However, it was claimed very soon
\cite{3r,4r,5r} that this phenomenon, which is also present in $pA$
collisions, could be due to the absorption of the pre-resonant $c\bar{c}$ pair
in the colliding nuclei. It is nowadays known \cite{6r,7r} that the data on
$pA$
and
on $AB$ collisions with a light projectile can indeed be described by nuclear
absorption with an absorptive cross-section $\sigma_{abs} = 7.3 \pm 0.6$ mb
\cite{6r}.

Recently the NA50 collaboration has found an anomalous $J/\psi$ suppression in
$PbPb$ collisions, i.e. a suppression which is substantially stronger than the
one obtained from nuclear absorption with the above value of $\sigma_{abs}$
\cite{7r,8r}. Two different interpretations of this anomalous suppression
have been
proposed in the literature. One is a conventional hadronic interpretation
in
which there is an extra $J/\psi$ suppression due to final state interaction of
the resonant $J/\psi$ state with co-moving hadrons \cite{9r,10r,11r,12r}.
The other interpretation \cite{6r,13r,14r,15r} assumes that when the
local hadronic energy density is larger than some critical value (taken to be
around the one reached in a central $SU$ collision), there is a discontinuity
in
the $J/\psi$ survival probability. (See also \cite{16r} for ideas based on
percolation of strings.)

\section{Nuclear absorption}
\hspace*{\parindent} Nuclear absorption is present in all interpretations. We
describe it in the probabilistic model of Ref. \cite{4r}. Let us consider first
proton-nucleus collisions. In this model, the pre-resonant $c\bar{c}$ pair is
produced at some point $z$ inside the nucleus and scatters with nuclei on its
path
at $z^\prime > z$, with an absorptive cross-section $\sigma_{abs}$. This
produces
a change in the $A$ dependence of the $J/\psi$ inclusive cross-section. For
nucleus-nucleus collisions this change is
given \cite{4r} by

\beq
S^{abs}(b, s) = {[1 - \exp (- A \ T_A(s) \ \sigma_{abs}) ] [1 - \exp (- B\
T_B(b-s) \ \sigma_{abs})] \over AB \ T_A(s) \ T_B(b-s) \sigma_{abs}^2}\
\ .
\label{1e}
\eeq

\noindent Here $T_A$ and $T_B$ are the nuclear profile functions, determined
from
a standard Saxon-Woods density, and $\sigma_{abs}$ is the absorptive
cross-section. In the following we take $\sigma_{abs} = 7.3 \pm 0.6$ mb which
gives the best fit to the $pA$ data \cite{6r}. Note that $S^{abs} = 1$ for
$\sigma_{abs} = 0$. Expression (\ref{1e}) has the meaning of a survival
probability of the $J/\psi$ due to nuclear absorption.

Since we are aiming at a quantitative analysis it should be emphasized that the
probabilistic formula, with its longitudinal ordering in $z$, can only be true
in
the low energy limit. Therefore it is important to evaluate the uncertainty
resulting from using this formula at $\sqrt{s} \sim 20$ GeV. In a recent paper
\cite{17r} the equivalent of Eq. (\ref{1e}) has been derived in a field
theoretical approach. The obtained formula is valid at all energies and
coincides exactly with (\ref{1e}) in the low energy limit. In the asymptotic
limit we find \cite{17r}
\beq
S_{\sqrt{s}
\to \infty}^{abs}(b, s) = \exp \left [ - {1 \over 2} \widetilde{\sigma} A
T_A(s) \right ] \exp \left [ - {1 \over 2} \widetilde{\sigma} B T_B(b - s)
\right ], \label{2e} \eeq

\noindent where $\widetilde{\sigma}$ is the $c\bar{c}-N$ total cross-section.
At
asymptotic energies, the results obtained from Eqs. (\ref{1e}) and (\ref{2e})
differ only by 8 $\%$. It is amazing that at $\sqrt{s} =$ 20 GeV the difference
between the result obtained with the exact formula, valid at all
energies, and the one obtained from
Eq.
(\ref{1e}) is less than 1 $\%$.

\section{Phase transition scenario}

\hspace*{\parindent} The anomalous $J/\psi$ suppression observed by the NA50
collaboration in $PbPb$ collisions came as a surprise. Indeed, it was believed
that the energy densities in $SU$ and $PbPb$ collisions were comparable - the
larger energies reached in $PbPb$ being compensated by a larger interaction
transverse area. However, it was soon realized that the local energy density
(i.e.
the energy per unit of transverse area $d^2s$) is larger by about 30 $\%$ in
central $PbPb$ than in central $SU$ collisions \cite{13r}. More precisely, let
us
consider the well known geometrical factors \cite{18r,19r}

\beq
m_{A(B)}(b, s) = A(B) \ T_{A(B)}(s) \left [ 1 - \exp \left ( - \sigma_{pp} \
B(A) \ T_{B(A)}(b-s) \right ) \right ]. \label{3e}
\eeq

\noindent The average number of participants is obtained as

\beq
\bar{n}_A + \bar{n}_B = {1 \over \sigma_{AB}} \int d^2b \int d^2 s \ N_w(b, s),
\label{4e}
\eeq

\noindent where

\beq
N_w(b, s) = m_A(b, s) + m_B(b, b - s).
\label{5e}
\eeq

\noindent In Ref. \cite{13r}, the quantity in Eq. (\ref{5e}) is taken as a
measure of the energy per unit transverse area at each impact parameter. It is
found \cite{13r} that the maximum value reached in $SU$ collisions at $b = 0$
is
$n_c =$ 3.3 fm$^{-2}$ (see Fig. 1). The model of Ref. \cite{13r} can then be
formulated as follows: For $N_w(b, s) < n_c$, one assumes nuclear absorption
alone with a standard value of $\sigma_{abs}$ (see Section 2). For $N_w(b, s)
\geq n_c$, one puts $\sigma_{abs} = \infty$, i.e. one assumes that none of the
$J/\psi$'s produced in this hot region can survive. This model gives the
maximal suppression that can be obtained in $PbPb$ collisions when one imposes
that, up to central $SU$ collisions, there is only nuclear absorption. The
interesting result is that with this simple model one approximately obtains the
suppression observed experimentally in central $PbPb$ collisions. In Ref.
\cite{6r} a more detailed analysis of the data has been performed in a similar
framework. More precisely, instead of the local energy density $N_w(b, s)$, the
authors of Ref. \cite{6r} consider the quantity

\beq
\kappa (b, s) = {N_c (b, s) \over N_w (b, s)}\ \ ,
\label{6e}
\eeq

\noindent where
\beq
N_c(b, s) = AB \ T_A(s) \ T_B(b - s) \sigma_{pp}\ \ .
\label{7e}
\eeq

\noindent The quantity (\ref{6e}) gives the average number of collisions per
participant at fixed $b$ and $s$. It increases with the centrality of the
collision.  The value of $\kappa (b, s)$ determines the onset of
deconfinement in Ref. \cite{6r}. More precisely, at a given impact parameter
$b$, the absorptive cross-section is assumed to be infinite for $\kappa (b, s)$
larger than some critical value $\kappa_c$, while for $\kappa (b, s) <
\kappa_c$ it
takes the standard value $\sigma_{abs} = 7.3 \pm 0.6$ mb discussed in Section
2.
Actually in Ref. \cite{6r} one has two distinct values of $\kappa_c$. One
is for $\chi$'s and is taken to be equal to the maximum value of $\kappa (b,
s)$ in a central $SU$ collision, i.e.

\beq
\kappa^{\chi}_c = \left [ \kappa (b = 0, s = 0) \right ]_{SU} \simeq 2.3\
\ .
\label{8e}
\eeq

\noindent The second one is for direct $J/\psi$ and is taken as a free
parameter satisfying $\kappa^{\psi}_c > \kappa_c^{\chi}$. The overall $J/\psi$
survival probability is then obtained by combining 40~$\%$ $\chi$ suppression
with 60~$\%$ suppression of direct $J/\psi$.

In this way a reasonable description of $J/\psi$ suppression is obtained.
The results for $PbPb$ are shown in Fig. 2.

Finally in Ref. \cite{14r} one combines nuclear absorption, deconfinement and
absorption by co-movers (see Section 4). The quantity which determines the
onset of deconfinement depends not only on $b$ and $s$ but also on the
longitudinal coordinate $z$. When some critical value of this quantity is
reached the involved cross-sections present a discontinuity but their values
remain finite.

\section{Absorption by co-moving hadrons}

\hspace*{\parindent} The survival probability of the $J/\psi$ due to absorption
with co-moving hadrons is given by (see \cite {6r,9r,10r,12r} and references
therein)

\beq
S^{co}(b, s) = \exp \left [- \sigma_{co} \ N_y^{co}(b, s)\   \ln
\left ( {N_y^{co}(b,
s) \over N_f} \right )\ \theta (N_y^{co}(b, s) - N_f)  \right
].  \label{9e}
\eeq

\noindent Here $N_y^{co}(b, s)$ is the density of hadrons per unit transverse
area $d^2s$ and
per unit rapidity at impact parameter $b$.
$N_f$ is the density at freeze-out that we take to be universal. The argument
of
the log
is the interaction time of the $J/\psi$ with co-moving hadrons. The
$\theta$-function is numerically irrelevant (see Section~5). $\sigma_{co}$ is
the
co-mover cross-section properly averaged over the momenta of the colliding
particles (the relative velocity of the latter is included in its definition).
We treat $\sigma_{co}$ as a free parameter \cite{20r,21r}. All species of
hadrons
are included in $N_y^{co}$. This quantity has been computed in the dual parton
model (DPM) \cite{22r}. It is expressed as a linear combination of the
geometrical quantities defined in Eqs. (\ref{5e}) and (\ref{7e}) \cite{12r}.

Note that
$S^{co}(b, s) = 1$ for $\sigma_{co}=0$. The effects of the co-movers in
proton-nucleus collisions turn out to be negligeably small (see Section 5).

The inclusive cross-section for $J/\psi$ production in nuclear collisions is
then given by

\beq
I_{AB}^{\psi}(b) = {I_{NN}^{\psi} \over \sigma_{pp}} \int d^2s \ N_c(b, s) \
S^{abs}(b, s) \ S^{co}(b, s),  \label{10e}
\eeq

\noindent where $N_c$ is given by (\ref{7e}). We see that for Drell-Yan pair
production ($\sigma_{abs} = \sigma_{co} = 0$), $I_{AB}^{DY} = I_{NN}^{DY}\
AB$. We take $\sigma_{pp}=30$ mb.

The results \cite{12r} are presented for three sets of parameters. Set I:
$\sigma_{abs} =$ 7.3~mb, $\sigma_{co} =$ 0; Set II: $\sigma_{abs} =$ 6.7 mb,
$\sigma_{co} =$ 0.6 mb, $N_f =$ 1.15 fm$^{-2}$; and Set III: $\sigma_{abs}
=$ 7.3 mb, $\sigma_{co}$ = 1.0 mb, $N_f =$ 2.5 fm$^{-2}$.

The results for $J/\psi$ suppression versus $AB$ are presented in Fig.~3. In
Fig.~4 we show the ratio $J/\psi$ over DY versus $E_T$ and in Fig.~5 we show
the same ratio versus $L$. This variable, defined in Refs. \cite{2r,8r}, is a
measure of the centrality of the collision. We see that nuclear absorption
alone (Set I) fails very badly, whereas Sets II and III give a reasonable
description of the data.

\section{Proton-nucleus collisions}

\hspace*{\parindent} Let us first discuss the role of the $\theta$-function in
Eq.
(\ref{9e}). This function looks like a discontinuity in the co-mover
contribution. However, numerically it has practically no effect in our results
for
$SU$ and $PbPb$ collisions - especially with the parameters of Set II. Indeed,
the
value $N_f = 1.15$ fm$^{-2}$ in this Set (which is the same used in Ref.
\cite{6r}) is just the average density of co-movers in $pp$ (given by $[1/(\pi
R_p^2)] dN/dy$). Therefore the argument of the log is always larger than one
(except for values of the integration variables where the argument of the
exponent
is very close to zero) and the $\theta$-function is irrelevant. With Eq.
(\ref{9e}), the effect of co-movers is rather small in $SU$ (see Figs. 3
and 4)
and
negligeable in $pA$.

In Ref. \cite{10r} a different expression for the argument of the log
was introduced. Using it in Eq.(\ref{9e}) for $AB$ collisions our results are
practically unchanged. However, the effect of the co-movers in $pA$, although
small, is then non-negligeable. When discussing the $\psi /DY$ ratio this
effect
can be compensated by a decrease of $\sigma_{abs}$. However, an effect is left
in
the ratio $\psi '/\psi$, which will decrease by roughly 10~$\%$ between $pp$
and
$pPb$.

\section{$\psi '$ suppression}

\hspace*{\parindent} There is a rather general consensus that (most of) the
observed $\psi '$ suppression in $SU$ and $PbPb$ collisions is due to co-mover
interactions. Even the authors \cite{6r}, who strongly support the deconfinement
scenario for $J/\psi$ suppression, do interpret in that way the $\psi '$ over
Drell-Yan ratio in all systems (including $PbPb$). Let me discuss this ratio
within Set~II. Here the value of $N_f$ is the same used in Ref. \cite{6r} and
the analysis therein is essentially unchanged in our approach. More precisely,
in $pA$ collisions this ratio is constant both in Ref. \cite{6r} and in our
formalism (Eq. (\ref{9e})). In $SU$ collisions one can choose the value of
$\sigma_{co}^{\psi '} > \sigma_{co}^{\psi}$ such as to reproduce \cite{6r,10r}
the
ratio $\psi '/DY$. In the formalism of Ref. \cite{6r} one then gets a good
description of $PbPb$ using this value of $\sigma_{co}^{\psi '}$. In our case,
the
value for the last bin of $PbPb$ is too small. However, this discrepancy is at
the
level of two standard deviations, and, therefore, one has to wait for better
data\footnote{In Ref. \protect{\cite{10r}} this discrepancy was taken seriously
and solved by introducing the exchange reactions $\psi + \pi \to \psi ' + X$
and
$\psi ' + \pi \to \psi + X$.}.

Let me discuss now the effect of deconfinement on $\psi '$ suppression. In
Ref. \cite{23r} it is shown that deconfinement produces a sharp decrease of
the ratio $\psi '/\psi$ at the critical value of the $\psi '$ deconfining
phase transitions (taken to be the same as for $\chi$'s, Eq. (\ref{8e})). At the
time Ref. \cite{23r} was published, it was assumed that this deconfinement
would take place after $pPb$. Thus, the decrease of the ratio $\psi '/\psi$
observed in $SU$ was regarded as a qualitative success of the phase transition
scenario. Nowadays, we know that this phase transition can only take place
after the last $E_T$ bin of $SU$ (Eq. (\ref{8e})). No sharp decrease of
the $\psi '/\psi$ ratio is observed here. Therefore the experimental behaviour
of
the ratio $\psi '/\psi$ (or $\psi '/DY$) does not confirm the qualitative
expectations of the phase transition scenario.

\section{The inverse kinematic experiment}

\hspace*{\parindent} Perturbative QCD calculations yield a very small value of
the
co-mover cross-section, i.e. of the cross-section of a bound $J/\psi$ with
hadrons near threshold \cite{20r}. When non-perturbative effects are
introduced, a
much larger value is obtained \cite{21r}. In order to settle this important
question Kharzeev and Satz have proposed to measure the $J/\psi$ suppression
in $pA$ collisions in the backward hemisphere (or $A$ on $p$ in the forward
one). When the $J/\psi$ is slow in the rest system of the nucleus it will be
produced inside the nucleus and if its interaction cross-section is very small
there will be no $J/\psi$ suppression. The problem, however, is that it may be
experimentally very difficult to distinguish a cross-section of, say, less
than 0.1 mb \cite{20r} from the moderate values (0.5 to 1.5 mb) needed in the
co-mover scenario.

Perhaps a more interesting possibility would be to consider $SU$ in the
backward hemisphere (or $U$ on $S$ in the forward one). Indeed, in the
co-mover scenario the decrease of the absorptive cross-section discussed
above may be over-compensated by an increase of absorption due to an increase
of the density of co-movers when one moves in rapidity towards the maximum of
$dN/dy$. The observation of a maximum in the $J/\psi$ suppression at this
value of $y$ would support the co-mover scenario. A further advantage of this
proposal is that it can be done in the rapidity range $(0 < y_{cm} < 1$)
covered by the present NA50 experimental set up. The calculation of this
effect is, of course, possible in the co-mover scenario but it requires a
hypothesis on the energy dependence of $\sigma_{abs}$. It should be stressed,
however, that the observation of such a maximum does not rule out the phase
transition scenario. Indeed, the $J/\psi$ environment at $- 1 < y_{cm} < 0$
in $SU$ (or at $0 < y_{cm} < 1$ in $US$) is hotter than the one in $SU$ at $0
< y_{cm} < 1$, and therefore deconfinement can be present there. This shows
that the geometrical criteria for deconfinement used in Refs. \cite{6r,13r}
can only be appropriate for a flat rapidity density of hadrons. When this is
not the case the critical value of the local energy density has to be a
function of $y$. It also shows the difficulty to disentangle the co-mover
scenario from the deconfining one.

\section{Discussion and conclusions}

\hspace*{\parindent} The main difference between color deconfinement and
co-mover
absorption is the presence in the former of a sudden change in the $J/\psi$
survival probability, whereas all changes are smooth in the latter.  However,
this discontinuity is assumed to take place in a variable (the local energy
density) which is not directly observable.

As for the number of free parameters, in the deconfinement approach of Ref.
\cite{13r} there is (apart from $\sigma_{abs}$ which is present in all
approaches)
only the value of $n_c$ (the critical value of the local energy density).
Moreover, one has chosen a value of $\sigma_{abs} = \infty$ above the critical
density. In Ref. \cite{6r} there is an extra parameter since one takes two
distinct critical values for $\chi$'s and for direct $J/\psi$
deconfinement.

In the co-mover scenario there are two free parameters: $\sigma_{co}$ and
$N_f$. The $E_T$ (or $L$) dependence in $PbPb$ is controlled by both
quantities. However, we see in Figs. 3, 4 and 5 that
changing $N_f$ by a factor 2
has very little effect on the results, since this change is compensated by a
corresponding change in the value of $\sigma_{co}$. Because the effects of
these two parameters are strongly correlated, it is not possible in the
simple co-mover approach of Refs. \cite{9r,10r,12r} to reproduce the
structure in the $E_T$ (or $L$) dependence of the $PbPb$ data. Taking a
mixture of $\chi$'s and direct $J/\psi$ with distinct parameters, as done in
Ref. \cite{6r}, would improve the situation. However, one should make sure
that such a structure is not due to a statistical fluctuation.

The success of the co-mover model is due to the fact that, when a couple
($N_f,\sigma_{co}$) is chosen in such a way to reproduce the $E_T$ (or
$L$) dependence in $PbPb$, the corresponding effect of the co-movers in $SU$
turns out to be rather small and does not spoil the success of the nuclear
absorption model. Moreover, the dependence on centrality of the local density
of co-movers is stronger in $PbPb$ than in $SU$ collisions. This produces the
change in the $L$ slope between the two systems, seen in Fig. 5.

In conclusion, to distinguish the deconfinement scenario involving a phase
transition (or discontinuity) from the co-mover scenario turns out to be a
quantitative rather than a qualitative issue. The present data can be
reasonably
well described in both approaches with a comparable (and small) number of free
parameters. So far no experiment has been suggested which would allow to
disentangle these two mechanisms. \par \vskip 5 truemm

\section*{Acknowledgements}

\hspace*{\parindent} 
We thank C. Gerschel and A. B. Kaidalov for discussions and
for
a stimulating collaboration (Ref. \cite{10r}), and D. Kharzeev, M.
Nardi
and H. Satz who pointed out to us a numerical error in the first version of
Ref.
\cite{12r}. N. A. also thanks the Xunta de Galicia for financial support.
\par 

\begin{figure}[hbt]
\vspace{6 truecm}
\caption{ The density $N_w(b, s)$ in Eq. (\protect{\ref{5e}}) for $s$
along the direction of the impact parameter, for various values of the impact
parameter $b = 0, 2, 4,...$ fm. The origin is at a distance $b/(1 + R_B/R_A)$
from the center of nucleus $A$. Left: $SU$ collision; right: $PbPb$
collision. The horizontal dashed line coresponds to the largest density
achieved in the $SU$ system, $n_c =$ 3.3 fm$^{-2}$.}
\end{figure}

\begin{figure}[hbt]
\vspace{6 truecm}
\caption{The experimental $J/\psi$ survival probability over $DY$
ratio \protect{\cite{7r}} divided by the deconfinement suppression in $PbPb$
collisions, with $\kappa_c^{\chi} =$ 2.3 and $\kappa_c^{\psi} =$ 2.9
\protect{\cite{6r}}.}
\end{figure}

\begin{figure}[hbt]
\begin{center}
\epsfig{file=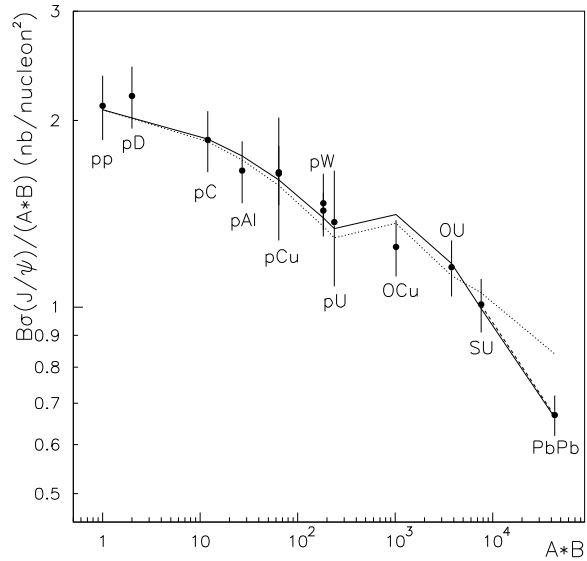,height=8.5cm}
\end{center}
\caption{$J/\psi$ suppression versus $AB$: Set I (dotted
line), Set II (solid line) and Set III (dashed line) compared to the
experimental data \protect{\cite{8r}}.
Note that the calculations have been
performed only for those nuclei where data exist. The obtained values have
been joined by straight lines. For clarity of the figure, the results of Set
III
are shown only for $SU$ and $PbPb$.}
\end{figure}

\begin{figure}[hbt]
\begin{center}
\epsfig{file=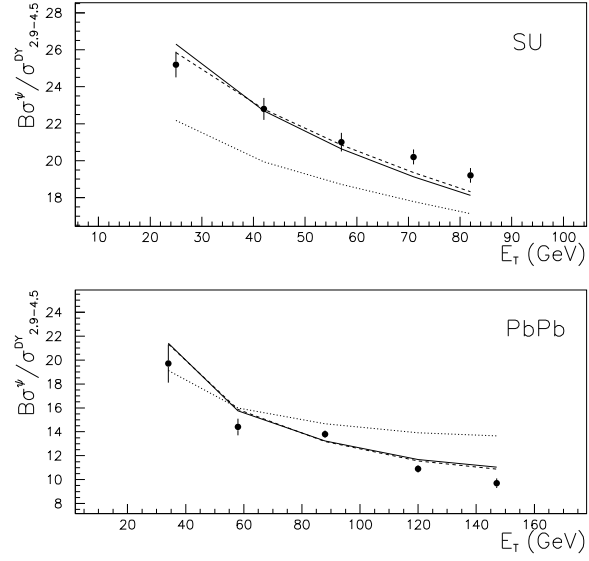,height=8.5cm}
\end{center}
\caption{$J/\psi$ over $DY$ ratio versus $E_T$: Set I (dotted
line), Set II (solid line) and Set III (dashed line) compared to the
experimental data \protect{\cite{8r}}.}
\end{figure}

\begin{figure}[hbt]
\begin{center}
\epsfig{file=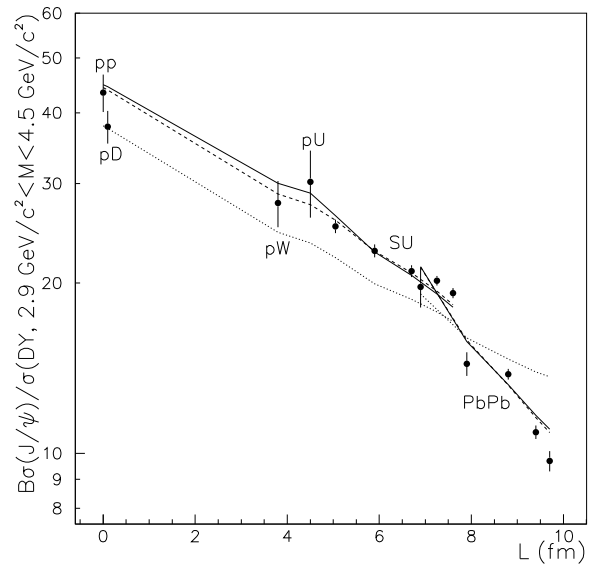,height=8.5cm}
\end{center}
\caption{Same as in Fig. 4 plotted versus $L$ \protect{\cite{2r,8r}}.}
\end{figure}

\end{document}